\documentclass[10pt,aps,longbibliography,rmp,twocolumn]{revtex4-1}
\usepackage{acronym,graphicx,hyperref}

\begin{document}

\title{Colloquium: Multimessenger astronomy with continuous gravitational waves
and future detectors}

\author{Benjamin J. Owen}
\affiliation{
Department of Physics, University of Maryland Baltimore County, 1000 Hilltop
Circle, Baltimore, Maryland 21250, USA
}

\begin{abstract}
Continuous gravitational waves from rapidly rotating neutron stars are on the
new frontiers of gravitational wave astrophysics and have strong connections to
electromagnetic astronomy, nuclear astrophysics, and condensed matter physics.
In this Colloquium I survey prospects for detection of continuous gravitational
waves from various neutron star populations, especially aided by electromagnetic
observations.
Although there are caveats, current theories and observations suggest that the
first detections are likely within a few years, and that many are likely in the
era of next generation detectors such as Cosmic Explorer and the Einstein
Telescope.
I also survey what can be learned from these signals, each one of which will
contain more cycles than all the compact binary mergers ever detected.
Since continuous gravitational wave emission mechanisms depend on aspects of
neutron star physics, such as crustal elasticity, which are not well constrained
by current astronomical observations and physical experiments, their detection
can tell us a great deal that is new about extreme matter.
Even more can be learned by combining gravitational wave observations with data
from the Square Kilometre Array, the Next Generation Very Large Array, FAST, and
other electromagnetic detectors operating in the next generation era.
\end{abstract}

\maketitle

\tableofcontents

\acrodef{O4}{fourth observing run}
\acrodef{O5}{fifth observing run}
\acrodef{O6}{sixth observing run}

\section{Introduction}

The discovery of gravitational waves from a binary black hole merger
\cite{LIGOBBH} began the era of gravitational wave astronomy.
The discovery of gravitational waves from a binary neutron star merger
\cite{LIGOBNS} accompanied by a variety of electromagnetic signals
\cite{LIGOMMA} began the era of multimessenger gravitational wave astronomy.
The LIGO-Virgo-KAGRA Collaboration continues to do ground breaking science on
more compact binary mergers \cite{O4aCatalog} with the ``Advanced'' generation
of detectors \cite{AdvancedLIGO, AVirgo, KAGRA} which completed their
\ac{O4} \cite{Capote2025, O4aIntro} in November 2025.

Even more breakthroughs \cite{Evans2023, Branchesi2023} are possible with next
generation gravitational wave detectors such as Cosmic Explorer \cite{Evans2021}
and the Einstein Telescope \cite{Punturo2010}.
Perhaps most tantalizing is the prospect of novel types of gravitational wave
signals such as continuous waves from long-lived asymmetries of non-merging
rapidly rotating neutron stars.
They have not been detected yet, but may be detected before long
\cite{Evans2023, Corsi2024, Gupta2024}.
Continuous waves are complicated signals from complicated sources tied to a
great variety of extreme physics and astrophysics.
Detecting them will be mathematically and computationally difficult, but can
reveal populations and effects very different from those involved in binary
mergers.

In this Colloquium I give an overview of continuous gravitational waves from
neutron stars, including projections of detectability and what can be learned
with detectors from the near future to the next generation.
Like all Colloquium articles, this one does not have as much detail as a true
review, but is meant to serve as an accessible introduction.
Throughout I will give pointers to more thorough reviews on various related
topics and will tend to cite key papers and recent papers containing surveys of
older work.
There have been a few next generation reviews and white papers including the
detectability of continuous waves~\cite{Kalogera2021, Evans2023, Branchesi2023,
Corsi2024, Gupta2024, ETScience} and one focusing on continuous waves
\cite{Jones2025}.
Here I include a broader survey and summarize the most recent detectability
estimates \cite{Owen2025}.
There is a longer history of reviews of what can be learned from a detection,
such as \textcite{Jones2021, Haskell2023, Jones2025} to mention some recent
ones.

For the rest of Section~I, I present background material.
In Sections~II and~III, I survey detection prospects and what can be learned
respectively.
In Section~IV, I summarize and look to the future.

\subsection{Improving detectors}

\begin{figure}
\includegraphics[width=0.48\textwidth]{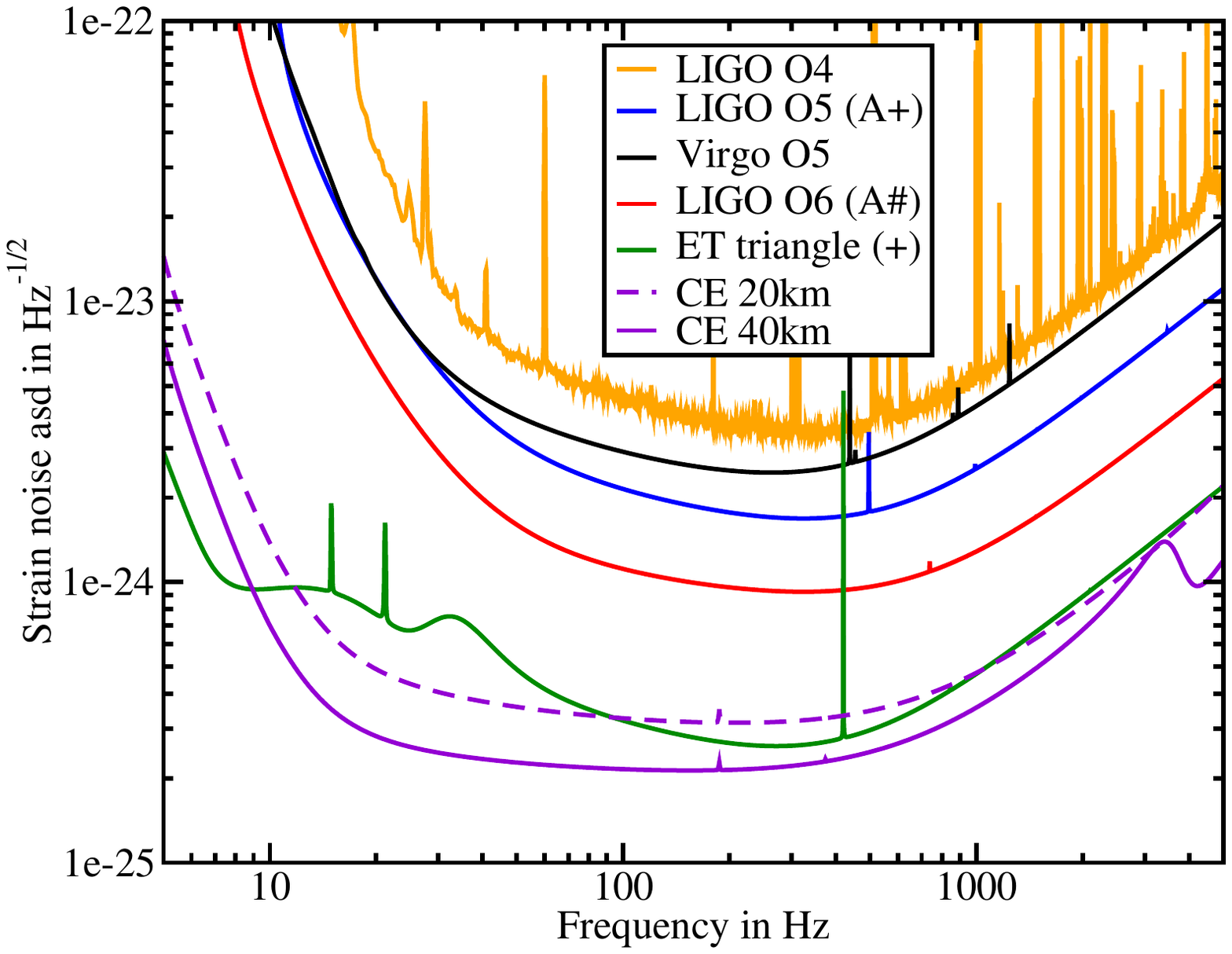}
\caption{
\label{fig:detectors}
Projected noise curves (one-sided strain noise amplitude spectral densities) of
some near future and next generation detectors.
For reference I have plotted the best current (O4) noise curve, from the LIGO
Livingston Observatory \protect\cite{Capote2025}.
}
\end{figure}

We can look forward to many improvements to ground-based gravitational wave
detectors over the next decade and beyond.
The LIGO-Virgo-KAGRA Collaboration finished its \ac{O4} run in November
\cite{Capote2025}.
Figure~\ref{fig:detectors} shows expected sensitivity improvements by plotting
noise curves---one sided strain amplitude spectral densities as functions of
frequency---for some detectors.
For reference I start with an \ac{O4} noise curve of the LIGO Livingston
Observatory \cite{Capote2025}.
The others include two near future versions of LIGO \cite{Aplus, Asharp}, one
near future version of Virgo \cite{Scenarios}, and the next generation Einstein
Telescope \cite{Hild2011} and Cosmic Explorer \cite{CE}.

More details on near future upgrades of present facilities can be found in
\textcite{Scenarios}.
The LIGO A+ upgrade is expected to begin the LVK Collaboration's \ac{O5} in a
few years~\cite{ObsURL}.
Virgo and KAGRA are expected to join \ac{O5} with their own improvements, though
with less sensitivity (more noise) than LIGO.
Virgo's \ac{O5} target noise curve is plotted in Fig.~\ref{fig:detectors};
KAGRA's noise curve is significantly higher and is not plotted.
The proposed LIGO A$^\sharp$ upgrade is expected to take data in the \ac{O6} a
few years later, and is likely the best that can be done with current
facilities.
In the {O6} era the two existing LIGO detectors in Hanford, Washington and
Livingston, Louisiana should be accompanied by a third in Aundha, India.

The Einstein Telescope and Cosmic Explorer are expected to be constructed in the
2030s, and involve new facilities due to their greatly increased arm lengths
which increase their gravitational wave strain sensitivities.
Both detectors will also benefit from improvements in optics, materials, and
seismic isolation.
Locations of these facilities are not yet finalized.
The Cosmic Explorer design \cite{Evans2021} is very similar to LIGO but ten
times larger, with 40\,km arms rather than 4\,km for approximately ten times
lower noise amplitude.
A 20\,km option is also being considered, either alone or in tandem with a
40\,km detector.
The Einstein Telescope default design is triangular, 10\,km rather than Virgo's
3\,km and putting two interferometers optimized for different frequencies in
each arm of the equilateral triangle \textcite{Hild2011}.
In Fig.~\ref{fig:detectors} I have, as usual, converted the Einstein Telescope
noise to an equivalent sensitivity for an L-shaped detector like the others
assuming a plus-polarized signal.
I have also, as usual, plotted the net noise curve for the high- and
low-frequency Einstein Telescope interferometers.

In later Figures I show joint sensitivity curves for various networks, or
combinations of the detectors mentioned above.
These are not quite the same set of networks treated in \textcite{Evans2023} and
\textcite{Gupta2024} because I am not showing some whose noise curves are
similar and create visually crowded plots.
But the basic idea is the same: to show a variety of possibilities for the near
future and the next generation era, including a third Advanced LIGO detector in
India and two possible versions of Cosmic Explorer.
Assuming that each detector takes a comparable amount (live time) of data, the
sensitivities add (in terms of noise power) harmonically (like resistors in
parallel).
That is, the network is dominated by the best one, or by the best ones if they
are approximately tied.
For continuous waves, unlike compact binary mergers and other short signals, the
network sensitivity depends relatively weakly on the locations of the source and
detectors since the detectors rotate around many times during the observation
\cite{Jaranowski1998}.
Thus, while Virgo and KAGRA are extremely helpful for detection and localization
of short signals like compact binary mergers \cite{LIGOBNS, Scenarios}, even
with higher noise curves, the sensitivity of the LIGO-Virgo-KAGRA network to
continuous waves is dominated by the LIGO detectors, especially at high
frequencies.
Therefore, for simplicity, I leave KAGRA off of the \ac{O5} and \ac{O6} curves
and Virgo off of the \ac{O6} curve.
For next generation detectors I focus on Cosmic Explorer plots related to recent
work \cite{Evans2023, Gupta2024, Corsi2024, Owen2025}, but note that the
Einstein Telescope has similar overall sensitivity to the 20\,km Cosmic
Explorer.

In parallel with gravitational wave detectors, electromagnetic observatories
will continue improving by leaps and bounds.
The Square Kilometre Array \cite{Dewdney2009} will grow from existing radio
telescope arrays in Australia and South Africa such as MeerKAT \cite{MeerKAT}.
It is expected to increase the number of detected pulsars, important targets for
continuous wave searches, from the few thousand known today \cite{ATNF} to about
20,000 \cite{Smits2009}.
The Next Generation Very Large Array will play a similar role in the United
States \cite{Wilner2024}.
The Five-hundred meter Aperture Spherical Telescope \cite{FAST} is already
operating in China and will undergo a series of significant upgrades.
More discoveries, not only of pulsars but of associated objects such as
supernova remnants, will arise from future high energy observatories such as
AXIS \cite{Reynolds2023} or COSI \cite{Tomsick2023}.
All of these will drive continuous wave multimessenger astronomy.

\subsection{Neutron stars}

Before we get into continuous gravitational waves, let's cover the basics of
neutron stars (which are the most likely sources).
This is a complicated topic, involving all ten volumes of the Landau and
Lifschitz series on theoretical physics, often taken to the greatest extremes of
matter in the modern universe.
For more thorough reviews see \textcite{Lattimer2021} for a nuclear perspective
on neutron star matter, \textcite{Caplan2017} for a condensed matter perspective
on it, \textcite{Yunes2022} for gravitational wave and electromagnetic
constraints on it, and \textcite{Ozel2016} for astronomical observations related
to it.
Since the latter is becoming dated, I recommend \textcite{Golomb2025} for an
update and summary of recent work.

Neutron stars are (probably) mostly made of neutrons, more than a solar mass of
them crammed into a ball with a radius about 12\,km \cite{Lattimer2021}.
This means an average density comparable to nuclear saturation density
($3\times10^{14}$\,g\,cm$^{-3}$) and a central density several times greater.
Such a density results from the collapsing core of a massive main sequence star
undergoing a supernova explosion.
Rising density during core collapse drives inverse beta decay, which neutronizes
the matter to a great extent, and the collapse is stopped before black hole
formation if neutron degeneracy pressure rises fast enough.
Calculations predict a maximum neutron star mass in the range 2--3\,$M_\odot,$
and observations support this, more likely on the low end of the range
\cite{Golomb2025}.
Since the core of a neutron star contains matter well above nuclear saturation
density, it might not consist of (mostly) neutrons but rather hyperons, meson
condensates, quarks, or various mixed phases.
In ``normal'' neutron star models there is a density dependent admixture of
protons, electrons, and (in most of the core) muons.
Neutrinos are constantly produced since the matter tends toward dynamic
beta-equilibrium, and emission of these neutrinos usually dominates cooling and
bulk viscosity \cite{Lattimer2021}.
In most neutron stars most of the matter should be in a superconducting
superfluid state, which strangely enough tends to increase rather than decrease
the viscosity through a complicated set of microphysical interactions
\cite{Mendell1991}.
The outer 1--2\,km of the neutron star consists of neutrons and protons in more
or less normal (although neutron rich) nuclei, acting macroscopically as a solid
crust which is of great interest for continuous gravitational waves.
Some models of exotic phases in the core also act like solids on macroscopic
scales \cite{Owen2005}.
Neutron star magnetic fields are expected to extend into the crust and be even
stronger than the observed external fields, but may not extend into the core
depending what type of superconductor the matter forms \cite{Lasky2015}.

More than three thousand neutron stars are currently known as pulsars
\cite{ATNF}.
That is, their electromagnetic emission is seen to pulse on and off regularly,
usually in radio but sometimes in gamma rays or x-rays or at other wavelengths.
While some details of the pulsar emission mechanism are not known, the gist of
it is clear \cite{Philippov2022}:
Radio waves are produced nonthermally by charged particles in the magnetosphere
and beamed to some extent, so that as the star rotates the beam intermittently
intersects the line of sight to Earth.
Clearly there must be many more neutron stars not seen as pulsars simply because
of unfortunate orientation; probably a factor 5--10 \cite{Tauris1998}.
Some x-ray pulsars may pulse due to thermal emission confined to hot spots on or
near the surface.
The period of pulsation tells us the spin frequency of the star, which can
exceed 700\,Hz \cite{ATNF}.
This is a factor of 2--3 lower than the maximum predicted by nuclear models of
the equation of state \cite{Lattimer2021}.
The gap is significant and it has been suggested that it may be due to
continuous gravitational wave emission (see below).

Pulsar spin frequencies typically drift downward on timescales of millennia or
longer, and may exhibit other small variations \cite{ATNF}.
The spin-down parameter or first time derivative of the frequency indicates that
these neutron stars have surface magnetic dipole fields exceeding $10^8$\,G,
often exceeding $10^{12}$\,G, and sometimes reaching $10^{15}$\,G
\cite{Philippov2022}.
Stars in the last category, called magnetars, spin too slowly to be of interest
as continuous gravitational wave sources.
Pulsar spin frequencies also exhibit Doppler modulation due to orbital velocity
if the pulsar is in a binary system, and some pulsars feature glitches or sudden
upward jumps in frequency.
This last, plus some remaining unknown frequency drifts called timing noise,
means that pulsars must be monitored frequently to obtain the spin phase which
is needed for the most sensitive gravitational wave searches \cite{ATNF}.

Many non-pulsing neutron stars are seen associated with supernova remnants
\cite{Green2025} or pulsar wind nebulae \cite{Safi-Harb2013}.

Almost all observations so far are related to the bulk numbers of mass, radius,
and spin \cite{Ozel2016, Yunes2022}.
That is because electromagnetic emission comes from the surface or
magnetosphere.
Observations of neutron star cooling give us some idea of electrical and thermal
conductivity and what layers are present in the interior.
Gravitational waves however couple directly to the high density matter in the
interior.
Continuous waves in particular couple to a great deal of microphysics such as
elasticity and viscosity.

\subsection{Continuous gravitational waves}

By now it is well known that gravitational waves are ripples in spacetime
predicted by Einstein's General Theory of Relativity, and many textbooks have
been written on the subject.
Specializing to continuous gravitational waves, this is still a large topic.
More detail on data analysis methods and (upper limit) results of past searches
can be found for instance in \textcite{Piccinni2022, Riles2023, Wette2023}.

Continuous gravitational waves are like pulsar signals: long lived, nearly
monochromatic, usually evolving (intrinsically) on a timescale of millennia or
more, most likely emitted by rapidly rotating neutron stars.
In one important way they are unlike pulsars---the gravitational waves are not
really beamed, though the strain amplitude varies by a bit more than a factor of
two as a function of orientation \cite{Riles2023}.
As the detectors rotate and move with the Earth, both gravitational wave
polarizations are observed even in L-shaped detectors.
(For short-lived signals such as compact binary mergers, an L-shaped detector
picks up only one polarization.)
Doppler shifts and polarization related modulations evolve on timescales of days
to years and carry information such as the sky location of the source
\cite{Jaranowski1998}.
In terms of mathematical sophistication and computational cost, continuous wave
searches are the most difficult type of gravitational wave search.

Although in this Colloquium I will mostly avoid the very complicated topic of
data analysis, it is helpful to understand some basics.
Because the signal model is well known (up to some parameters) in advance,
continuous gravitational waves are amenable to matched filtering and similar
methods---in the limit where infinite computing power is available.
Originally developed for radar during World War Two, matched filtering, which is
essentially integration of a set of template waveforms against the data, is the
most sensitive analysis technique and is commonly used with compact binary
mergers \cite{Riles2023}.
It and similar techniques can be sensitive to changes of order a radian in the
signal phase, and therefore can require many templates to cover the signal
parameter space.
Because a continuous wave signal is ``on'' during an entire observing run of a
year or more, even modest searches can easily use $10^{13}$ templates or more
and a million core-hours on a modern computing cluster.
Large searches such as all sky surveys use distributed computing and are still
limited to coherent integration times of a few days, and combine these short
stretches semi-coherently to cover longer data sets \cite{Wette2023}.

Searches for continuous gravitational waves can be aided greatly by information
from electromagnetic astronomy, which can narrow the search parameter space
considerably.
Based on that information, continuous gravitational wave searches can be divided
into four types of differing computational cost and sensitivity \cite{Owen2009}.
Each type has many examples, of which I cite a few (mostly recent ones).
Known pulsar searches like \textcite{LIGOCrab, O4aPulsars}, which use a sky
position and full timing solution (usually from radio telescopes, sometimes from
a high energy satellite) are the cheapest and most sensitive because they can
precisely target a single point in signal parameter space (frequency, Doppler
modulation, etc).
If they target $r$-modes (quasinormal modes described below) they search a
broader band but still do well \cite{Fesik2020, Rajbhandari2021,
LIGOJ0537rmodes}.
With sky positions but no timing data, directed searches for non-pulsing neutron
stars which are accreting from a non-compact companion \cite{LIGOScoX1,
Whelan2023, LIGOMarkov, Vargas2025} or not \cite{CasAVelaJr, Ming2025, Owen2024}
have intermediate cost and sensitivity.
Small area searches \cite{LIGOGalacticCenter} are similar.
X-ray variability of accreting neutron stars implies a fluctuating torque and
spin frequency, which is an additional complication \cite{LIGOMarkov,
Vargas2025}.
All-sky all-frequency surveys \cite{LIGOAllSky}, sometimes checking for
potential binary orbits \cite{Covas2022, Covas2025}, are also performed.
But due to the huge parameter space they have the highest cost and least
sensitivity.

Continuous wave searches are intrinsically multimessenger.
Depending on the target, they may use pulsar timing data, sky positions of point
sources or small regions, or x-ray variability data from non-pulsing accreting
neutron stars.
By the time of next generation gravitational wave detectors, the Square
Kilometre Array \cite{Dewdney2009} is expected to bring the number of known
pulsars from a few thousand to about twenty thousand \cite{Smits2009}.
Future x-ray satellites such as AXIS will contribute more pulsars, better
observations of accreting neutron stars, and more point sources in supernova
remnants \cite{Safi-Harb2023}.
Even gravitational-only discoveries from all-sky gravitational wave surveys
should result in electromagnetic detections, since analysis of the signal will
yield a precise sky position and frequency evolution (see below).
And any such discovery will certainly become a high priority target for radio
telescopes and high-energy astronomy satellites.

Continuous gravitational wave emission mechanisms from neutron stars are
complicated, potentially relating to all the aspects of neutron star physics
mentioned above.
Good surveys are given by \textcite{Lasky2015} and \textcite{Glampedakis2018},
which also summarize other neutron star gravitational wave signals yet to be
discovered.
Here let's cover the basics.

The most straightforward emission mechanism is a ``mountain'' on a rotating
neutron star---more strictly, a mass quadrupole (spherical harmonic $\ell=2,$
$|m|=2$ or 1) that rotates with the star.
Much of the continuous wave literature expresses the mass quadrupole in terms of
a dimensionless ellipticity,
\begin{equation}
\epsilon = \left( I_{xx} - I_{yy} \right) / I_{zz},
\end{equation}
where $I_{ab}$ is the moment of inertia tensor and the $z$-axis is the spin
axis.
This is comparable to the mass quadrupole over the principal moment of inertia,
or the height of a mountain in units of stellar radii \cite{Owen2005}.

Like terrestrial mountains, neutron star mountains can be supported by elastic
stresses in the solid crust \cite[and references therein]{Morales2022}.
If some more exotic particle physics models are right, ``mountains'' could be
buried in the core \cite{Owen2005}.
Most of the literature deals with the maximum mass quadrupole produced by a
mountain, which can be written schematically as
\begin{equation}
\mathrm{(breaking\ strain)} \times \mathrm{(shear\ modulus)} \times
\mathrm{(geometry)}
\end{equation}
integrated over the star.
At the high pressures ($10^{34}$\,dynes\,cm$^{-3}$) found inside neutron stars,
the breaking strain of any crystal is probably close to the ideal value of order
$10^{-1}$ \cite{Horowitz2009}---and an order of magnitude higher than the best
terrestrial materials.
The geometry factor depends on the structure of the star, including how much of
it is solid.
Maximum elastically supported ellipticities for normal neutron stars are likely
more than $10^{-6}$ \cite{Morales2022}.
But that just tells us what ellipticity a star \textit{could} have, not what it
\textit{should} have.
The latter depends on the history of the star and it is poorly known what would
drive it toward the maximum except in accreting stars (see below).

Unlike terrestrial mountains, neutron star mountains can be supported by
magnetic stresses too \cite[e.g.]{Melatos2005, Lander2014}.
The matter is a good conductor and will avoid crossing field lines.
A dipole magnetic field tends to produce a mass quadrupole
\cite[e.g.]{Melatos2005, Lander2014}.
The precise value of the quadrupole for a given (average) $\mathbf{B}$ depends
on the field configuration and the presence or absence of superconductivity.
All but the very youngest neutron stars, mostly neglected here, have cooled
enough to have superconducting protons.
There are many models of the internal magnetic fields in this case.
Roughly speaking the field is probably poloidal dominated with a toroidal
component in the crust, forming a ``twisted torus.''
This \textit{should} support an ellipticity a few times
\begin{equation}
\label{emagnetic}
\epsilon \sim 10^{-8} B / 10^{12}\,\mathrm{G,}
\end{equation}
where $B$ is the average internal field \cite{Lander2014}.

The other main emission mechanism considered here is a long-lived mass-current
quadrupole due to an $r$-mode sustained by the Chandrasekhar-Friedman-Schutz
instability, a gravitational wave driven version of the well known two stream
instability \cite{Andersson1998, Friedman1998}.
Depending on viscosity and other damping mechanisms in neutron star matter,
$r$-modes may last thousands of years in young neutron stars if they are
spinning fast enough \cite{Bondarescu2009}.
$R$-modes may be revived in older neutron stars being spun up by accretion
\cite{Bildsten1998, Andersson1999} and continue for millions of years after
accretion stops \cite{Reisenegger2003}.
Much of the literature uses an $r$-mode amplitude $\alpha$ \cite{Lindblom1998}
defined as roughly $\delta v$ in units of the (uniform) background velocity of
the fluid due to stellar spin, and I will use that too.
The instability means that $\alpha$ will grow (if the spin and temperature are
right) from any initial perturbation, and presumably will saturate at some
amplitude which is most commonly thought to be determined by nonlinear
hydrodynamics.
The highest estimates of the $r$-mode saturation amplitude using perturbative
mode-mode coupling are of order $10^{-3}$ \cite{Bondarescu2009}.

I will point out but skip over several interesting but uncommon, speculative, or
slightly non-continuous emission mechanisms.
Free precession \cite{Jones2001} of a lumpy neutron star can lead to continuous
wave emission, but it is probably rarely detectable.
In recent years many groups have noticed that dark matter candidates such as
axion clouds around black holes could lead to continuous wave signals, and there
are already a couple of reviews on this topic \cite{Piccinni2022, Miller}.
Dark matter might also interact directly with gravitational wave detectors,
producing something similar to a continuous wave signal \cite{KAGRADarkMatter}.
Pulsar glitches may be associated with long transients, lasting many days and
therefore benefiting from continuous wave data analysis techniques even if they
are not truly continuous waves \cite[and references therein]{LongTransients}.

\section{Detectability}

First, let us be clear that we \textit{could} detect continuous gravitational
waves at any time.
An all sky survey could get lucky with one of the many neutron stars undetected
by electromagnetic astronomy.
For a known pulsar with observed spin-down parameters a key milestone is the
\textit{spin-down limit.}
This is the intrinsic strain implied if all of the observed spin-down is due to
gravitational wave emission and is given for ``mountain'' emission by
\cite[e.g.]{Owen2025}
\begin{equation}
h_0^\mathrm{sd} = 8.06\times10^{-19} \left( \frac{\mathrm{1\,kpc}} {D} \right)
\left( \dot{P} \right)^{1/2} \left( \frac{\mathrm{1\,s}} {P} \right)^{1/2},
\end{equation}
where $D$ is the distance to the pulsar and $P$ and $\dot{P}$ are the spin
period and its first time derivative.
(It is slightly different for $r$-modes.)
Obviously this is an upper limit on what can be expected.
For continuous waves, since the detector strain response is amplitude modulated
daily by the changing orientation of the detectors, it is convenient to talk
about an \textit{intrinsic strain} $h_0$.
This is, roughly speaking, the gravitational wave amplitude with all the source
and detector angles factored out---see \textcite{Jaranowski1998} for the (gory)
details.
Pulsar searches first beat a spin-down limit in \textcite{LIGOCrab} [or
\textcite{Rajbhandari2021} for $r$-modes] and continue to do so more and more
\cite{O4aPulsars}.
Especially at lower frequencies, next generation ground-based detectors and
deci-Hz space-based detectors will beat spin-down limits for hundreds of pulsars
\cite{Pagliaro2025}.

But it is more interesting to ask when we \textit{should} detect something.
That is, without invoking a star strained to its maximum mountain height by some
unknown mechanism, or quarks or other exotica in the core, an improbably large
magnetic field, or an emission mechanism with many uncertainties like the
$r$-modes.
And \textit{with} an argument for a minimum amplitude.
Below we shall focus especially on two cases of \textit{should}, and briefly
survey some (perhaps less likely) cases of \textit{could.}

It is convenient to make detectability estimates using the \textit{sensitivity
depth}
\begin{equation}
\mathcal{D} = \sqrt{S_h} / h_0,
\end{equation}
which is intended to give an idea of how well a search does when the detector
noise is factored out.
Here $S_h$ is the noise power spectral density (the square of the amplitude
spectral density), and should be taken as the harmonic mean over detectors,
times, etc.
Here we are talking about upper limits on it, typically at 90\% or 95\%
confidence.
Sensitivity depth has been estimated for practically all past continuous wave
searches \cite{Wette2023}, so we have a good idea of what is possible.
Taking it too literally however is dangerous, since it depends on the amount
or live time of data available as well as the computing time and method used.
For a simple coherent integration of a single pulsar timing solution over a
total one year of data (distributed between any number of similarly sensitive
interferometers) $\mathcal{D}$ is about 500\,Hz$^{-1/2}$ and increases as the
square root of the data time available.
For an all-sky search with tougher statistics due to sampling many points in
parameter space, and with shorter coherent integration times, $\mathcal{D}$ can
be a few tens and scale as the fourth root of data time \cite{Wette2023}.

Because I am focusing more on should than could, below I make conservative
estimates based on previous observing runs, which were generally shorter than O4
and planned future runs.
Keep in mind that with long runs and stable detectors, sensitivity depths
could be significantly better.

\subsection{Accreting neutron stars}

\begin{figure}
\includegraphics[width=0.48\textwidth]{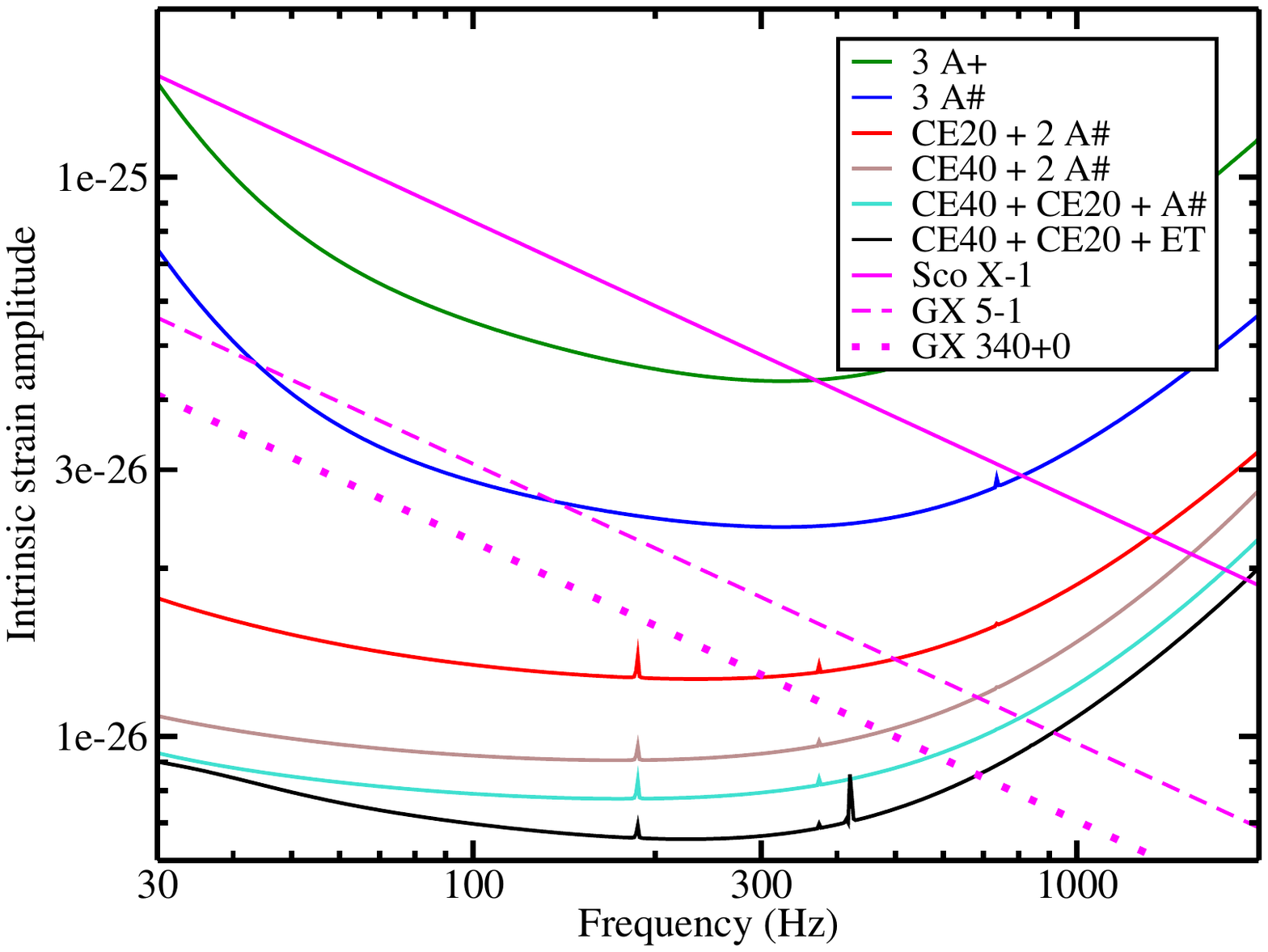}
\caption{
\label{fig:lmxbs}
Projected detectability of the most rapidly accreting neutron stars.
The slanted lines (Sco~X-1 \textit{etc.}) show the intrinsic strain $h_0$ as a
function of the unknown spin frequency for the brightest few sources, assuming
that accretion torque is balanced by gravitational wave emission.
The curves show the sensitivities of various detector networks, assuming a
conservative sensitivity depth of 39\,Hz$^{-1/2}.$
Adapted from \protect\textcite{Owen2025}.
}
\end{figure}

Accretion, being nonaxisymmetric, is a natural way to generate the rotating
quadrupoles needed for continuous gravitational wave emission.
The star's magnetic field should channel the infalling plasma somewhat, and the
observation of x-ray pulses in some accreting neutron stars is evidence of this.
If accretion creates a large enough lateral temperature gradient in the crust,
it can lead to asymmetric electron capture layers, producing a series of
``mountains'' due to changing chemical composition buried deep in the crust
\cite{Bildsten1998}.
Neutron star magnetic fields also may confine accreted plasma on the surface
well enough to produce significant crustal ``mountains'' held in place by
magnetic stresses \cite{Melatos2005}.
And accretion heating and spin-up may drive the $r$-modes of old neutron stars
unstable again \cite{Bildsten1998, Andersson1999}.

There is indirect evidence that such gravitational wave emission is happening.
The steep frequency dependence of gravitational wave emission has been proposed
as a natural way of regulating the spins of accreting neutron stars: 
Accretion spins up a star until the gravitational wave spin-down torque balances
the spin-up torque from accretion \cite{Papaloizou1978b, Bildsten1998}.
Maybe the star cycles around equilibrium \cite{Levin1999}, especially since
accretion rates fluctuate.
This means that overall the loudest gravitational wave sources should be the
brightest in x-rays.
That is good for getting a sky position to guide the gravitational wave search.
It is not good for getting spin frequencies, which for the brightest sources are
unknown and likely to wander with varying accretion rate.
These sources are known in the astronomy literature as the Z-source population
of low mass x-ray binaries \cite{Watts2008}.

Figure~\ref{fig:lmxbs} shows the detectability of various accreting neutron
stars assuming torque balance.
The curved lines are sensitivities for various detector networks assuming a
sensitivity depth of 39\,Hz$^{-1/2}$---a conservative choice for recent searches
\cite{Wette2023}.
The straight lines in Fig.~\ref{fig:lmxbs} are torque balance limits (functions
of the unknown gravitational wave frequency) for the top few sources using
average x-ray fluxes from \textcite{Watts2008}.
Several more would lie just below the lines shown, but I have left them off
since the plot becomes crowded.

Assumptions here are generally conservative:
The signal lines could go up (improve) by tens of percent with different
assumptions \cite{LIGOScoX1} and the sensitivity curves could go down (improve)
by tens of percent assuming a depth achieved by some comparable searches
\cite{Wette2023}.
So the signal curves could lie almost a factor of two further above the noise
than plotted here if we are very lucky.

Sco~X-1 is obviously the best, being the brightest extrasolar source in the
x-ray sky.
It is borderline detectable already even with spin wandering if it is at low
frequency \cite{Vargas2025}.
By the time of A$^\sharp$, GX~5\textminus1 becomes detectable similarly, and
Sco~X-1 become detectable at gravitational wave frequencies up to 1\,kHz.
By the time of the full next generation network, Sco~X-1 is all but certain to
be detected if torque balance holds, since a 2\,kHz gravitational wave frequency
is well above what is possible for the fastest known pulsar \cite{ATNF}.
And several others are likely to be detected.
Conversely, even if none of these is detected by the next generation era, that
would yield a major payoff by ruling out the venerable torque balance theory.

If X-ray observations find pulsations and thus a spin period for one of these
objects, that will improve its sensitivity depth and curve by up to an order of
magnitude \cite{Wette2023}.
If the torque balance argument holds, this would probably result in a detection
of Sco~X-1 even with near future detectors.
Therefore continuing attempts to detect pulsations with present and future X-ray
satellites \cite{Galaudage2021} are important to multi-messenger astronomy.
Refinements to the orbital elements can help too---for example, recent searches
\cite{LIGOScoX1, LIGOMarkov} were improved \cite{Whelan2023, Vargas2025} by
using corrected ephemerides with smaller uncertainties \cite{Killestein2023}.
This correction and improvement were due to switching from observations of the
neutron star and disc (where spectral lines are broad and fluctuating) to the
narrow lines emitted by the irradiated hot spot on the companion, illustrating
that improved analysis can make significant gains even with existing
observatories.

\subsection{Millisecond pulsars}

\begin{figure}
\includegraphics[width=0.48\textwidth]{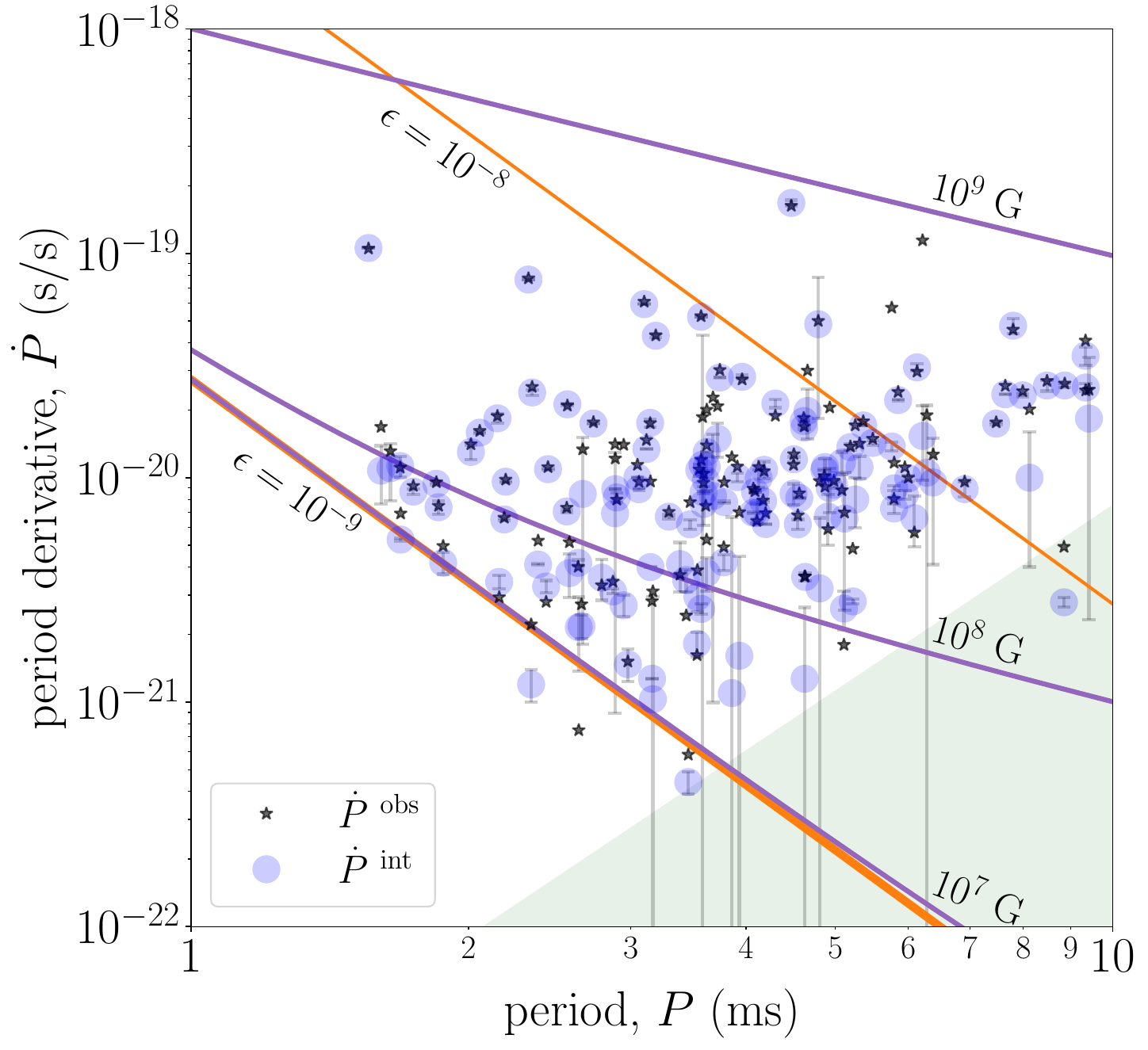}
\caption{
\label{fig:woan}
Spin periods and period derivatives of millisecond pulsars, showing evidence for
a quadrupole cutoff corresponding to a minimum ellipticity of about $10^{-9}.$
Reproduced from \protect\textcite{Woan2018} under
\href{https://creativecommons.org/licenses/by/3.0/} {Creative Commons
Attribution license 3.0}.
}
\end{figure}

Millisecond pulsars, so called because they have spin periods on the order of
milliseconds, are interesting targets because there is observational evidence
and a theoretical argument that they \textit{should} have a minimum ellipticity,
and that that ellipticity is high enough to be interesting.

Astronomers commonly plot the locations of pulsars in the $P$-$\dot{P}$ diagram,
where $P$ is the spin period and $\dot{P}$ is its time derivative.
If $\dot{P}$ is dominated by magnetic dipole spin-down, pulsars of a given
surface magnetic field will lie along a straight line in the log-log plot
($\dot{P} \propto P^{-5}$).
\textcite{Woan2018} point out that, if one considers only millisecond pulsars,
there is a statistically significant cutoff corresponding instead to a minimum
quadrupole spin-down ($\dot{P} \propto P^{-7}$), as shown in
Fig.~\ref{fig:woan}.

A minimum quadrupole is difficult to explain with magnetically dominated
spin-down, which should be dipole dominated, but is natural if there is a
minimum mass quadrupole corresponding to about a $10^{-9}$ ellipticity
\cite{Woan2018}.
Such a quadrupole can be produced by magnetic stresses due to an internal
magnetic field of order $10^{11}$\,G, which is much stronger than the external
fields (inferred from $\dot{P}$) of order $10^8$\,G but weaker than the fields
of many young pulsars.
For many years the standard story of millisecond pulsar formation has been
\cite[e.g.]{Bhattacharya1991}:
Take a young pulsar with a surface field of order $10^{11}$\,G, accrete plasma
on it, spin the star up and bury most of the field with $10^8$\,G extending
outside to produce the observed $\dot{P}.$
In hindsight, this story naturally produces a minimum ellipticity of order
$10^{-9}.$
The argument is not airtight---for instance, the internal magnetic field might
decay on timescales more like millions of years \cite{Pons2019} than billions
\cite{Goldreich1992}---but it neatly dovetails with the observations
\cite{Woan2018}.

\begin{figure}
\includegraphics[width=0.48\textwidth]{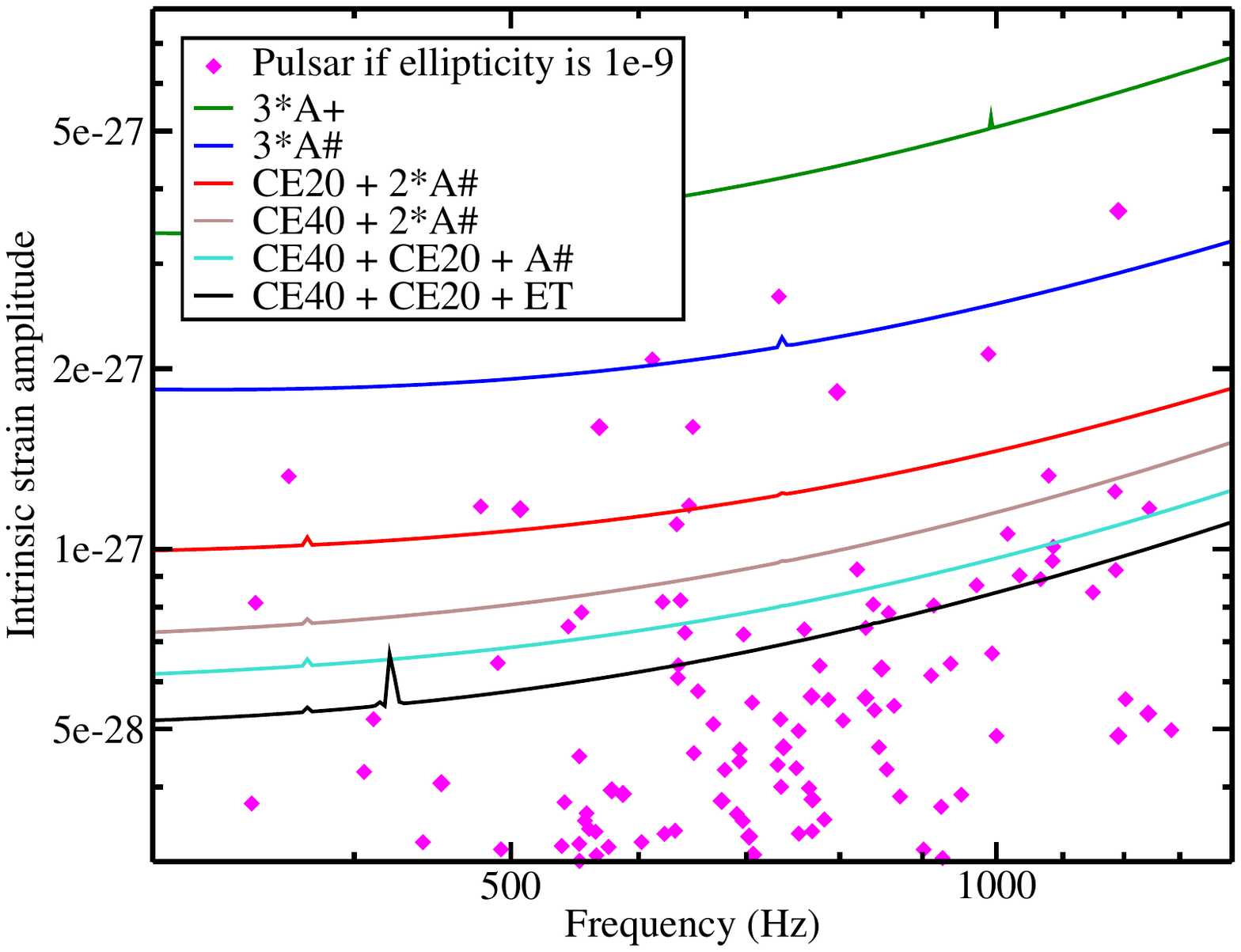}
\caption{
\label{fig:msps}
Projected detectability of currently known millisecond pulsars.
The sensitivity depth is assumed to be 500\,Hz$^{-1/2}.$
Diamonds show intrinsic strain amplitudes of known pulsars assuming mass
quadrupole emission at twice the spin frequency with an elliptcity of $10^{-9}$
or the spin-down limit, whichever is lower.
Adapted from \protect\textcite{Owen2025}.
}
\end{figure}

In Fig.~\ref{fig:msps} we see the implications of millisecond pulsars having a
$10^{-9}$ ellipticity.
The curves show projected sensitivities of searches with various detector
networks assuming a sensitivity depth of 500\,Hz$^{-1/2}.$
The diamonds correspond to intrinsic strains of known pulsars from version 2.5.1
of the ATNF catalog \cite{ATNF} assuming an ellipticity of $10^{-9}$ or the
spin-down limit if that is known and lower (it is in a handful of cases).
Some recent pulsar searches such as \textcite{O4aPulsars} are already close to
$10^{-9}$ ellipticity.

Keep in mind that these numbers are rough at the factor of two level, and are
fairly conservative.
The sensitivity depths of some pulsar searches have been twice as good as
assumed here \cite{Wette2023}.
Observationally, the ellipticity cutoff is not really a step function at
$10^{-9}$ \cite{Woan2018}.
Theoretically, for a given internal magnetic field the mass quadrupole can vary
by a factor of a few depending on mass and configuration.
And since pulsars certainly have different external fields, the internal fields
are likely to vary too.
Also, the strongest internal magnetic fields may decay with a time constant of
order a million years \cite{Pons2019}.

Nevertheless a clear picture emerges that several pulsars are likely detectable
with upgrades of existing facilities (A$^\sharp$ or comparable) and dozens
should be detectable in the era of next generation detectors.
This is without even counting the new pulsars to be found with the Square
Kilometre Array and x-ray satellites.
Many of the best pulsars in Fig.~\ref{fig:msps} have been found recently,
including the best one \cite{Clark2023}.
It is important to keep monitoring these pulsars, which are not always the ones
radio and high-energy astronomers focus on.
As we shall see below, even one detection can tell us a great deal.
And with Cosmic Explorer and the Einstein Telescope, even with no detections we
would get the interesting consolation prize of drastically revising the standard
theory of millisecond pulsar formation.

\subsection{Other populations}

While accreting pulsars and millisecond pulsars may be the most promising
in terms of specific objects that \textit{should} be detected, there are also
interesting prospects for young pulsars, electromagnetic point sources, and
all-sky surveys that \textit{could} be detected.
That is, their maximum ellipticities are consistent with detection, but we do
not know if they will be near their maxima.
And $r$-modes \textit{could} be detected from all types of source.
While an instability drives $r$-modes toward their maximum amplitude, there is
much uncertainty about what range of temperatures and spins allow the
instability to operate \cite{Lasky2015, Glampedakis2018}.

First, what if we repeat the above analyses of accreting neutron stars and
millisecond pulsars assuming $r$-mode emission rather than mountain emission?
For $\alpha = 10^{-3}$ and a sensitivity depth of 100\,Hz$^{-1/2}$
\cite{Wette2023} the plot looks qualitatively similar to Fig.~\ref{fig:msps}
\cite{Owen2025}.
Several pulsars are detectable, if they have that $r$-mode amplitude, by next
generation detectors and perhaps even with three A$^\sharp$ detectors.
However these are millisecond pulsars rather than young pulsars, and $r$-modes
would evolve differently in this population.
While there is an argument for active $r$-modes in some millisecond pulsars
\cite{Gusakov2014}, the neutron stars would probably be hotter than what is
observed in these pulsars as much of the $r$-mode energy would be dissipated as
heat \cite{Schwenzer2017}.
Also the saturation amplitude is probably up to a hundred times lower in these
stars and therefore they are more promising candidates for all sky surveys for
nearby objects---see \textcite{Covas2022} for such a search and more discussion
of the theoretical ups and downs.
One topic that commonly arises in discussions of $r$-modes is the fact that the
saturation amplitude calculations are difficult and delicate, and have not been
performed by a variety of groups and methods; and so one might still hope for a
higher amplitude and it is worth looking.

For some young known pulsars the $r$-mode situation might be more promising.
The predicted saturation amplitude is relatively high \cite{Bondarescu2009}.
For one pulsar, known by its celestial coordinates J0537\textminus6910, there
are hints from x-ray timing observations that its spin-down might be dominated
by gravitational waves from $r$-modes \textcite{Andersson2018}.
$R$-mode pulsar searches have been performed only in recent years and have
targeted this pulsar and the well known Crab pulsar \cite{Fesik2020,
Rajbhandari2021, LIGOJ0537rmodes}.

What about mountains on young known pulsars?
Since young pulsars generally spin down rapidly, they \textit{could} generally
have ellipticities of $10^{-6}$ or more consistent with their observed
$\dot{P}.$
It is not clear if they \textit{should} in the same way that \textcite{Woan2018}
argue for old millisecond pulsars.
Neutron stars are stably stratified, so there is no analog of plate tectonics.
But supernova explosions, which can kick newborn pulsars at speeds exceeding
1000\,km/s \cite{ATNF}, are certainly asymmetric; and one can argue in general
terms that some of that asymmetry might be reflected in the structure of young
neutron stars.
Also there is a subtle argument that long term timing of the Crab pulsar implies
an internal magnetic field producing an ellipticity of order $10^{-6}$
\cite{DallOsso2017}.
The results of plotting something like Fig.~\ref{fig:msps} for $10^{-6}$
ellipticity (or the spin-down limit if that is worse) are fairly predictable for
young pulsars \cite{Owen2025}.
The Crab and Vela pulsars, arguably the two most famous young ones, are
detectable at this ellipticity with any network containing a next generation
detector, and maybe even with three A$^\sharp$ detectors.
Surprisingly, more than a dozen nearby millisecond pulsars have spin-downs
(barely) consistent with an ellipticity of $10^{-6},$ and those could be
detectable on next generation timescales.
However some of the spin-downs may be contaminated, for example by binary
interactions, and in general there are not arguments for why millisecond pulsars
(which are an old population) \textit{should} have such high ellipticities, as
opposed to the $10^{-9}$ minimum from \textcite{Woan2018}.
These are still interesting to search, but it is harder to argue that they must
yield a detection.

Supernova remnants and point sources associated with them are another population
of relatively young neutron stars, generally being highly visible for only a few
tens of thousands of years \cite{Safi-Harb2013}.
Given the scalings of gravitational wave emission, the best targets might be the
youngest in our galaxy, the closest (which is also young), and the youngest in a
nearby galaxy---see \textcite{CasAVelaJr, Ming2025, Owen2024} for recent
searches and references to earlier work.
Extrapolating from these results, next generation detectors might detect
normal neutron star ellipticities or $r$-modes with amplitude $10^{-3}$ over a
wide frequency band in nearby neutron stars, and in the remnant of Supernova
1987A above a few hundred~Hz \cite{Owen2025}.
Many supernova remnants and point sources are interesting targets, but one must
keep in mind that many of the neutron stars will spin too slowly for Earth-based
detectors \cite{ATNF}.

Last but not least, all sky surveys:
Here we have to rely on population synthesis models such as
\textcite{Pagliaro2023} and references therein.
Our galaxy should contain $10^8$ or more neutron stars, vastly more than those
visible as pulsars or in supernova remnants and similar structures.
Population synthesis suggests that about one of these previously unknown neutron
stars could radiate continuous gravitational waves detectable with three
A$^\sharp$ level detectors, and the number could be of order one hundred in the
era of Cosmic Explorer and the Einstein Telescope.

\section{What can be learned from a detection}

A continuous wave signal contains a great deal of information.
Even one will have $10^{9}$--$10^{11}$ cycles after a year of observation, more
than all the binary mergers detected in our lifetimes.
Every six months the Earth orbits $3\times10^8$\,km through the wave field,
implying a diffraction limited position resolution of order an arcesecond
\cite{Jaranowski1999}.
In contrast to compact binaries, this is possible even with one interferometer.
The Earth-orbit baseline is even enough to yield a parallax distance if a source
is within a few hundred pc \cite{Sieniawska2023}.
The frequency and spin-down parameters (time derivatives of the frequency) will
be measured to 7--8 significant figures \cite{Jaranowski1999}.
And once a source is known it can be monitored with future observing runs,
making its parameters ever better known.
Even more information can be obtained by comparing the gravitational wave signal
with its accompanying electromagnetic signal.

\subsection{Frequency ratio}

The most accurately measurable parameter is the gravitational wave frequency
\cite{Jaranowski1999}.
Its ratio to the spin frequency obtained from electromagnetic observations tells
us the gravitational wave emission mechanism.
To summarize a slightly complicated flowchart \cite{Jones2021}:
In most cases we should easily discriminate between free precession (multiple
harmonics), a mountain (most likely a ratio of or near 2), and $r$-modes (a
ratio around 1.5).
Gravitational wave polarization also is different between mountains and
$r$-modes, although it will not be as precisely measured \cite{Owen2010}.

A frequency ratio of 2 is most likely coming from a mountain.
If the ratio is not quite 2, that tells us in addition that the crust and core
are rotating slightly out of sync \cite{LIGOCrab}, perhaps leading up to a
glitch.
The difference tells us the crust-core coupling timescale, which is likely
related to viscous timescales in the liquid core.
In an accreting neutron star, magnetically supported mountains might be
indicated by cyclotron resonant scattering features in the x-ray emission
\cite{Haskell2015}, allowing us to distinguish between them and elastically
supported mountains such as those from buried electron capture layers.

A frequency ratio of about 1.5 means $r$-modes.
If we detect $r$-modes, we know that dissipative processes such as viscosity are
not too large \cite{Owen1998}.
The precise $r$-mode frequency ratio is about 1.39--1.64 \cite{Idrisy2015,
Ghosh2023} and depends mainly on the compactness $GM/Rc^2$, where $M$ is the
star's mass and $R$ is its radius.
Hence a frequency measurement yields information on the bulk properties, which
in turn constrain the equation of state, a real prize for nuclear
astrophysicists.
Most neutron star models predict a roughly universal $R$ independent of $M$
\cite{Lattimer2021}, and if this is true it will tell us the mass of the star.
If the neutron star is in a binary, that can yield an independent mass
measurement.

\subsection{Distance}

Distances to continuous wave sources can be measured electromagnetically by
various means.
Radio pulses exhibit dispersion as they propagate through the interstellar
plasma, and via measurements and models of the galactic distribution the
dispersion measure can yield a distance estimate which is usually good to tens
of percent \cite{ATNF}.
Supernova remnant distances can be measured by various means, primarily the
kinematic method of matching the Doppler shift of interstellar hydrogen to the
velocity profile of the Galaxy \cite{Safi-Harb2013}.
Very long baseline interferometry can yield sky position measurements of point
sources accurate enough to get parallax distances for sources several kpc away
\cite{ATNF}.
A distance measurement allows a gravitational wave amplitude measured on Earth
to be converted into the amplitude of the quadrupole on the neutron star
\cite{Sieniawska2021}.
If we are fortunate enough to find a source within a few hundred~pc, we may get
a parallax directly from the gravitational waves \cite{Sieniawska2023}.

The quadrupole can tell us much.
Above we noted that electromagnetic observations will tell us if we are dealing
with a mass quadrupole (mountain) or current quadrupole ($r$-mode).
Long observations of the spin-down parameters (see below) are needed to
distinguish between elastically and magnetically supported mountains.
If we know which one we are dealing with, we get a probe of the interior of the
star.

Elastically supported mountains are maybe easiest to understand.
The shear modulus goes up with pressure and density \cite{Owen2005}.
Normal neutron stars are solid only in the crust, so the shear modulus and thus
the maximum quadrupole are relatively low, with a maximum ellipticity  of order
$10^{-5}$--$10^{-6}$ \cite{Morales2022}.
Speculative models such as baryon-quark hybrids and quark stars involve matter
deep in the core with higher density and shear modulus, and therefore can
support elastic quadrupoles several orders of magnitude higher \cite{Owen2005}.
Therefore observation of a large elastic mountain indicates an exotic
composition.

Magnetically supported mountains have implications for the (average) magnetic
field inside the star.
For most stars the field is expected to be a linear function of ellipticity as
in Eq.~\ref{emagnetic}, so a measurement of ellipticity yields the average
internal field \cite{Lander2014}.
Surprisingly, a detection strong enough to yield multiple peaks in the spectrum
could constrain the overall internal magnetic field configuration
\cite{Lasky2013}.

In the case of a current quadrupole from an $r$-mode, the saturation amplitude
is tied to dissipative mechanisms inside the star such as viscosity
\cite{Lindblom1998, Bondarescu2009}.
Viscous damping may be much higher in stars with various exotic phases such as
hyperons \cite{Lindblom2002, Ofengeim2019}.
Thus an $r$-mode amplitude observation could yield information not only on the
equation of state of a neutron star but on its composition.
Combined with equation of state information from the frequency \cite{Idrisy2015,
Ghosh2023} the distance allows us to disambiguate the star's moment of inertia
and $\alpha$ \cite{Ghosh2023b, Annamalai2024}.

\subsection{Spin-down parameters}

The spin-down parameters, especially higher order ones obtained over long
observation baselines, can yield further insights into the gravitational wave
emission mechanism of each source.
Together the frequency and first derivative yield information on the equation of
state ($M$ and $R$ separately) if magnetic field effects are not too large
\cite{Caride2019}.
Year-long observations, with the distance known, can separate out the dipole and
quadrupole (magnetic and gravitational wave) parts of the spin-down
\cite{Lu2023, Hua2023}.
This can be used to make a case for the mass quadrupole being elastically or
magnetically supported, since the external magnetic dipole field is likely
correlated with the internal field contributing to the mass quadrupole.

\section{Summary and outlook}

The take-away messages are simple.

Continuous gravitational waves are an exciting discovery frontier in the coming
years.
According to our current understanding of accreting neutron stars and their
descendants the  millisecond pulsars, these sources are likely to be detected
with upgrades of existing detector facilities.
In the era of next generation detectors such as Cosmic Explorer and the Einstein
Telescope, there should be many detections---and if not, it drastically revises
our last few decades of understanding the life cycles of these objects.

Even one detection will yield a great deal of information on neutron stars at
the extremes of condensed matter physics, nuclear physics, and several other
areas.
And this information will be qualitatively different from the first tantalizing
hints we are now learning from multimessenger neutron star binary mergers---it
will tie in to elasticity, viscosity, and other properties of the microphysics
of extreme matter, not just the bulk equation of state.
Gravitational waves opened a new window on the universe, and continuous
gravitational waves will peer through that window into the hidden depths of
neutron stars and the most extreme states of matter.

\acknowledgments

This work was supported by NSF grant PHY-2450793 to UMBC.
I am grateful to Alessandra Corsi and Michael Kr\"amer for discussions about
current and future radio observations and to Ian Jones for discussions about
gravitational waves.
Thanks also to Matthew Pitkin and Graham Woan for discussions about and
permission to use their figure.
This material is based in part upon work supported by NSF’s LIGO Laboratory
which is a major facility fully funded by the National Science Foundation.

\bibliography{cw}

\end{document}